# ROLE OF OXYGEN AND CARBON IMPURITIES IN THE RADIATION RESISTANCE OF SILICON DETECTORS

S. Lazanu [1], I. Lazanu [2]

[1] National Institute for Materials Physics,
POBox MG-11, Bucharest-Magurele, Romania

[2] University of Bucharest, Faculty of Physics,
Bucharest-Magurele, POBox MG-11, Romania

**Abstract**

**The influence of oxygen and carbon impurities on the concentrations of defects in silicon for detector uses, in complex fields of radiation (proton cosmic field at low orbits around the Earth, at Large Hadron Collider and at the next generation of accelerators as Super-LHC) is investigated in the frame of the quantitative model developed previously by the authors. The generation rate of primary defects is calculated starting from the projectile - silicon interaction and from recoil energy redistribution in the lattice. The mechanisms of formation of complex defects are explicitly analysed. Vacancy-interstitial annihilation, interstitial and vacancy migration to sinks, divacancy, vacancy and interstitial impurity complex formation and decomposition are considered. Oxygen and carbon impurities present in silicon could monitor the concentration of all stable defects, due to their interaction with vacancies and interstitials. Their role in the mechanisms of formation and decomposition of the following stable defects: $VP$, $VO$, $V_2$, $V_2O$, $C_i$, $C_iO_i$ and $C_iC_s$ is studied. The model predictions could be a useful clue in obtaining harder materials for detectors at the new generation of accelerators, for space missions or for industrial applications.**

## *1. Introduction*

Silicon-based devices have found applications in a wide variety of hostile radiation environments. For this reason, radiation effects in silicon represent a research field that have received extensive attention, in order to assess the performances of silicon to radiation. In the next future, silicon detectors will be used for many applications in particle or astroparticle physics as, for example, at the CERN Large Hadron Collider (LHC) and Super-LHC (S-LHC) or in space mission as, e.g. the Anti-Matter Spectrometer (AMS). The detectors used in these environments will be exposed long-time and/or to high fluences of charged and neutral particles. Various systematic studies have been performed to understand better the origins and consequences of radiation damage in silicon detectors. These cover the production of primary defects [1], their annealing mechanisms [2], [3], [4], [5], [6], [7], [8], the correlation with the characteristics of the irradiation particle [9], [10], [11] and with initial material impurities [12], [13]. The microscopic modifications of the material characteristics produce changes in the detector parameters. The microscopic phenomena and their consequences at the device level are still poorly understood.

In the last decade a lot of studies have been done to investigate the influence of different impurities, especially oxygen and carbon, as possible ways to enhance the radiation hardness of silicon for detectors in the future generation of experiments in high energy physics - see, e.g. references [14] and [15]. These impurities added to the silicon bulk modify the formation of electrically active defects, thus control the macroscopic device parameters. The model developed previously by the authors [16], [17], [18] and used in the present paper confirms these conclusions. The correlation established by the model between the initial material parameters, the rate of defect generation in the continuous irradiation regime, the production of defects and their time annealing, could be a useful clue in obtaining harder materials for detectors.

## 2. Macroscopic damage in silicon

Most semiconductor radiation detectors are based on the properties of the p-n diode junction. During irradiation, due to the collisions of the incident particle with the nuclei placed in the sites of the lattice, as well as due to the re-arrangement of the corresponding recoils, vacancies and interstitials, primary radiation defects, are generated. Consequences of irradiation processes, some characteristics of the devices are modified because of the formation of secondary defects. Some of them are electrically active, with energy levels located in the forbidden band gap, that act either as capture centres, or as generation – recombination ones. The principal effects are: bulk material resistivity modification, change of depletion voltage, increase of leakage current, change of electric field distribution in irradiated silicon p-n junction, decrease of the collection efficiency and decrease of the signal to noise ratio. [19], [20], [21].

Silicon used in high energy physics detectors is n-type high resistivity (1 ÷ 6 KΩcm) phosphorus doped FZ material. The concentrations of interstitial oxygen and substitutional carbon in silicon are strongly dependent on the growth technique. In high purity Float Zone Si, the concentrations of interstitial oxygen are around $10^{15}$ cm$^{-3}$, while in the oxygenation technique developed at BNL, an interstitial oxygen concentration of the order $5 \times 10^{17}$ cm$^{-3}$ is obtained. These materials can be enriched in substitutional carbon.

## 3. Radiation environments considered in the present analysis

In the present paper, two types of applications. In the first class of applications, the radiation field produced by cosmic rays is considered, for orbits in the neighbourhood of the Earth, at an altitude of about 380 km, where the dominant particles are the protons. Considering the flux spectra measured by the Alpha Magnetic Spectrometer (AMS) during space shuttle flight STS-91F [22], a differential generation rate of vacancy-interstitial pairs of $2 \times 10^2$ VI pairs/cm$^3$/s has been calculated [23]. The energetic differential generation rate has been calculated as the integral of the convoluted spectrum of the measured flux and of the energy dependence of the concentration of primary defects (CPD) for proton irradiation.

The second application is represented by the radiation field in the region of tracker detectors at the LHC accelerator, as well as to the new generation of accelerators, as S-LHC. Without loss of generality, the radiation field simulated for the CMS silicon tracker geometry [24] is considered in the following calculations. The spectra of charged hadrons (pions, kaons and protons) simulated for the positions of the silicon layers are taken from reference [24]. The hadrons are predominantly low-energy charged pions and protons, which are present in different amounts and have different energy spectra as a function of the distance in respect to the interaction point, and of the pseudorapidity. In all cases, the pions are the dominant particles. The energetic differential generation rates of defects have been calculated similarly as in the first case. The range of generation rates (considering the extreme positions in the tracker detector) are: $6.2 \times 10^8$ VI pairs/cm$^3$/s for pions and $5.6 \times 10^7$ VI pairs/cm$^3$/s for protons, and $8.1 \times 10^6$ VI pairs/cm$^3$/s and $3.1 \times 10^6$ VI pairs/cm$^3$/s for pions and protons respectively. This represents nearly seven orders of magnitude higher generation rate in respect to the field of the cosmic rays in the near Earth orbit. A ten fold increase in LHC design luminosity (at S-LHC [25]) is planned, reflected in an increase of the dose in the central region, and corresponding to a ten times increase of the generation rate of primary defects.

## 4. Modelling of radiation effects

The primary incident particle, having kinetic energy with values in the intermediate up to high energy range, interacts with the semiconductor material. After this process, the recoil nuclei resulting from these interactions lose their energy in the lattice. Their energy partition between displacements and ionisation is considered in accord with the Lindhard theory [26] and authors' contributions [27] and after this step the concentration of primary defects is calculated. The basic assumption of the present model is that the primary defects, vacancies and interstitials, are produced in equal quantities and are uniformly distributed in the material bulk. They are produced by the incoming particle, as a consequence of the subsequent collisions of the primary recoil in the lattice, or thermally. The generation term ($G$) is the sum of two components: $G_R$ accounting for the generation by irradiation, and $G_T$, for thermal generation. The concentration of the primary radiation induced defects per unit fluence (CPD) in silicon has been calculated as the sum of the concentrations of defects resulting from all interaction processes, and all characteristic mechanisms corresponding to each interaction process, using the explicit formula from reference [27] (see also concrete relations and details). Due to the important weight of annealing processes, as well as to their very short time scale, CPD is not a direct measurable physical quantity.

In silicon, vacancies and interstitials are essentially unstable and interact via migration, recombination, and annihilation or produce other defects. In some previous papers (see, e.g. refs. [16], [17], [18]), the authors developed a quantitative phenomenological model to explain the production of primary defects and their time evolution toward stable defects, starting from silicon with different quantities of initial impurities, considering different rates of primary defects production and conditions of irradiation. Without free parameters, the model is able to predict the absolute values of the concentrations of defects and their time evolution toward stable defects.

In the chemical reaction description, the formation and evolution of defects in silicon, around room temperature, could be described as follows:

| | | |
|---|---|---|
| $V + I \underset{G}{\overset{K_1}{\rightleftarrows}} 0$ (1a) | | $V + V \underset{K_5}{\overset{K_4}{\rightleftarrows}} V_2$ (2a) |
| $I \overset{K_2}{\rightarrow} \text{sinks}$ (1b) | | $I + V_2 \overset{K_6}{\rightarrow} V$ (2b) |
| $V \overset{K_3}{\rightarrow} \text{sinks}$ (1c) | | |
| $V + P \underset{K_7}{\overset{K_4}{\rightleftarrows}} VP$ (3a) | $V + O \underset{K_9}{\overset{K_4}{\rightleftarrows}} VO$ (4a) | $I + C_s \overset{K_1}{\rightarrow} C_i$ (5a) |
| $I + VP \overset{K_8}{\rightarrow} P$ (3b) | $I + VO \overset{K_{10}}{\rightarrow} O_i$ (4b) | $V + C_i \overset{K_{11}}{\rightarrow} C_s$ (5b) |
| $V + VO \underset{K_{12}}{\overset{K_4}{\rightleftarrows}} V_2O$ (6a) | $I + V_2O \overset{K_{13}}{\rightarrow} VO$ (6b) | $C_i + C_s \underset{K_{15}}{\overset{K_{14}}{\rightleftarrows}} C_iC_s$ (6c) $\quad C_i + O_i \underset{K_{16}}{\overset{K_{14}}{\rightleftarrows}} C_iO_i$ (6d) |

The reaction constants $K_i$ (i = 1, 4 ÷ 16) have the general form: $K_i = C \cdot v \cdot \exp(-E_i/k_BT)$, with $v$ the vibration frequency of the lattice, $E_i$ the associated activation energy and $C$ a numerical constant that accounts for the symmetry of the defect in the lattice. The reaction constant related to the migration of interstitials and vacancies to sinks could be expressed as: $K_j = \alpha_j \cdot v \cdot \lambda^2 \cdot \exp(-E_j/k_BT)$, with j = 2 (interstitials) and 3 (vacancies), $\alpha_j$:- the sink concentration and $\lambda$ - the jump distance.

The values of the activation energies are from the literature see for example reference [18].

The reactions are grouped into four categories: the first is related only to primary defects and their reciprocal interactions (equations 1 and 2); the second involves the reactions of primary defects with impurities, as well as complex decomposition (equations 3, 4 and 5), while the interactions of complexes, and of complexes with primary defects are comprised into the group of reactions represented by equations 6. Complexes of other impurities with primary defects could also be added to this reaction scheme.

In the analysis which follows, four types of Oxygen and Carbon doping have been considered for high resistivity silicon ($10^{14}$ cm$^{-3}$ Phosphorus): $2 \times 10^{15}$ O/cm$^3$, $3 \times 10^{15}$ C/cm$^3$; $4 \times 10^{17}$ O/cm$^3$, $3 \times 10^{15}$ C/cm$^3$; $2 \times 10^{15}$ O/cm$^3$, $6 \times 10^{17}$ C/cm$^3$; and $4 \times 10^{17}$ O/cm$^3$, $6 \times 10^{17}$ C/cm$^3$.

In Figure 1, the time evolution of the primary defects in the conditions of 10 years of continuous irradiation specified before is presented. The time evolutions of the concentrations of complexes defects: $V_2$, VP and $V_2O$, relevant for modifications of device parameters are represented in the same conditions (doping and irradiation) in Figure 2.

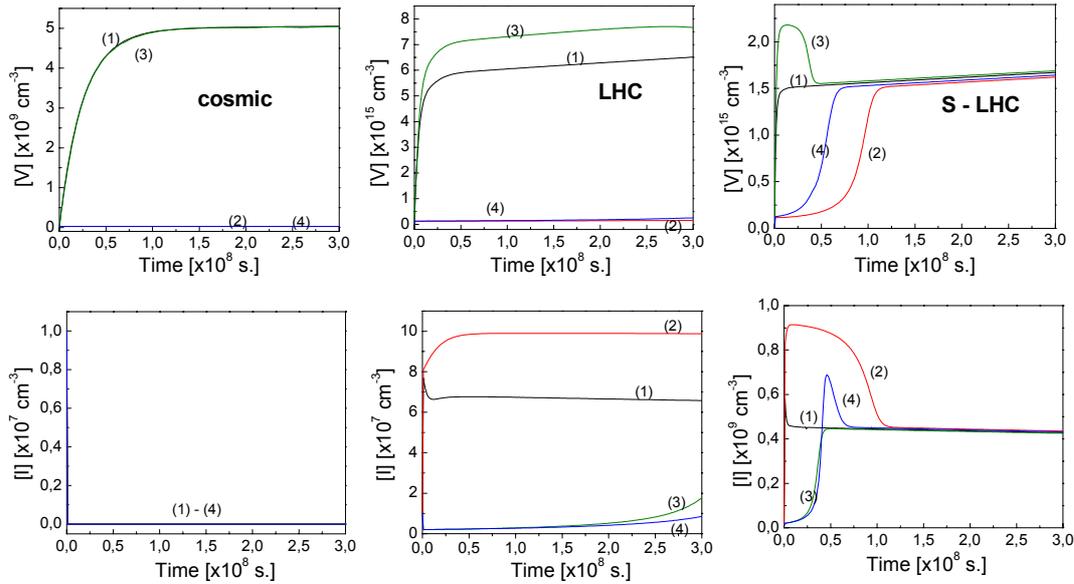

Figure 1
Time dependence of the concentrations of (top to bottom): vacancies and interstitials in conditions of continuous irradiation (left to right) cosmic, LHC and Super-LHC, for silicon with $10^{14}$ cm$^{-3}$ P, with the following concentrations of Oxygen and Carbon: $2 \times 10^{15}$ O/cm$^3$, $3 \times 10^{15}$ C/cm$^3$ - curve (1), $4 \times 10^{17}$ O/cm$^3$, $3 \times 10^{15}$ C/cm$^3$ - curve (2), $2 \times 10^{15}$ O/cm$^3$, $6 \times 10^{17}$ C/cm$^3$ - curve (3), $4 \times 10^{17}$ O/cm$^3$, $6 \times 10^{17}$ C/cm$^3$ - curve (4).

At low rates of generation of primary defects, due to the greater mobility of interstitials in respect to vacancies, the interstitial concentration remains constant in time, and independent on the concentration of impurities. In the extreme case of S-LHC conditions, before about three years of continuous irradiation, for materials with high concentration of oxygen, some possible processes initiated by interstitials are favoured, due to the relatively high concentration of interstitials present in the sample. After longer times, the peculiarities of different materials are lost and the concentrations of vacancies and interstitials remain practically constant in time. For an order of magnitude lower generation rate of primary defects, the particularities of material doping are relevant.

$C_i$, is highly mobile at room temperature and $C_s$ can be considered immobile. By a kick-out mechanism [28], substitutional carbon can be transformed into an interstitial defect. Interstitial oxygen acts as a sink for vacancies, thus reducing the probability of formation of complexes, associated with deeper levels inside the gap. Oxygen addition produces a dramatic decrease of vacancy concentration in all generation conditions, because oxygen fixes vacancies in the form of VO and $V_2O$ complexes. The most pronounced is this effect in lean carbon silicon (curve 2). In what regards the concentration of interstitials, oxygen addition keeps a higher concentration in the sample (curves 2 and 4) because more vacancies are fixed by oxygen, favouring this way the formation of interstitial – impurity complexes. Due to the competitive processes, the effects are different for different generation rates, being the most pronounced for S-LHC conditions. A remarkable point is that in these conditions, after approximately three years of operation, both interstitial and vacancy concentrations attain a plateau.

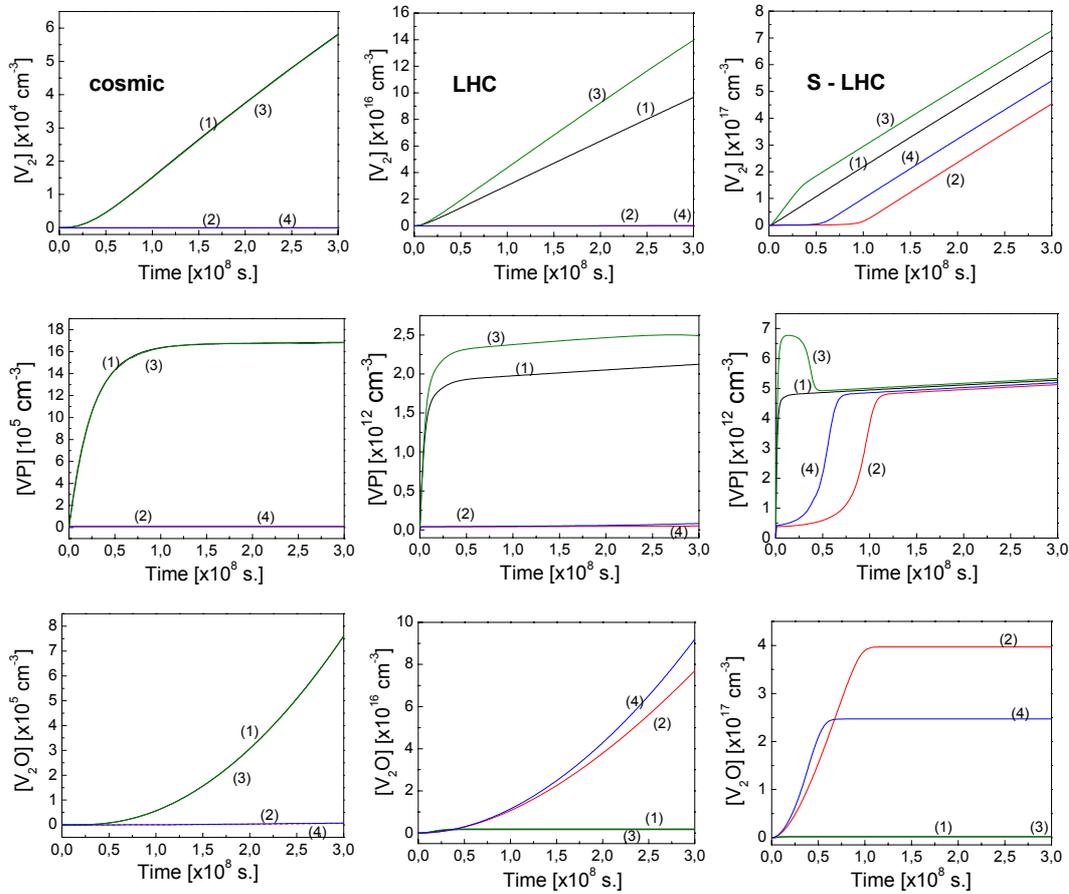

Figure 2
Time dependence of the concentrations of (top to bottom): $V_2$, VP and $V_2O$ defects in conditions of continuous irradiation for ten years for (left to right): cosmic, LHC and Super-LHC, for silicon with $10^{14}$ cm$^{-3}$ P, with the following concentrations of Oxygen and Carbon: $2\times10^{15}$ O/cm$^3$, $3\times10^{15}$ C/cm$^3$ - curve (1), $4\times10^{17}$ O/cm$^3$, $3\times10^{15}$ C/cm$^3$ - curve (2), $2\times10^{15}$ O/cm$^3$, $6\times10^{17}$ C/cm$^3$ - curve (3), $4\times10^{17}$ O/cm$^3$, $6\times10^{17}$ C/cm$^3$ - curve (4).

In conditions of cosmic irradiation, a clear distinction is seen between silicon with low concentration of oxygen (curves (1) and (3)), and silicon –rich material (curves (2) and (4)), while the concentration of carbon does not matter. At much higher irradiation rates (LHC and S-LHC), both the content of oxygen and of carbon is very important: For the LHC case, the highest concentration of divacacies and vacancy-phosphorous complexes are found in low oxygen, high carbon concentration material, correlated with the high concentration of vacancies calculated (see Figure 1), while for the $V_2O$ complex, oxygen enrichment is essential (curves (2) and (4)). At even higher radiation rates, the hierarchy is maintained. The remarkable point here is that V, I, VP and $V_2O$ concentration are saturating for all types of materials studied, all of them after the same amount of time for the same material, while all the generated vacancies are found in the increase of the $V_2$ concentration.

## *Conclusions*

In the frame of an original phenomenological model developed previously by the authors, the role of oxygen and carbon impurities in the mechanisms of formation and the kinetics of the following stable defects: $VP$, $VO$, $V_2$, $V_2O$, $C_i$, $C_iO_i$ and $C_iC_s$ in silicon for detector uses, in complex fields of radiation has been investigated. The obtained results, presented and discussed in the paper, could be a useful clue in obtaining harder materials for detectors at the new generation of accelerators, for space missions or for industrial applications.